\documentclass[twoside,twocolumn,10pt]{article}

\usepackage{fancyhdr}
\usepackage{geometry}
\usepackage{graphicx}
\graphicspath{{figures/}{../figures/}{./}}
\usepackage{titlesec}
\usepackage{ragged2e}
\usepackage{amssymb}
\usepackage{amsmath}
\usepackage{mathptmx}
\usepackage{booktabs}
\usepackage{array}
\usepackage[backend=biber,style=ieee]{biblatex}
\usepackage[font=footnotesize,labelfont=bf]{caption}
\usepackage{float}
\usepackage{indentfirst}
\fancypagestyle{firstpage}{
  \fancyhf{}
  \fancyfoot[L]{\footnotesize © 2026 Yasu. Released under CC-BY 4.0.}
  \fancyfoot[R]{\footnotesize August 2026}

}

\titleformat{\section}{\normalsize\bfseries}{\thesection.}{0.6em}{}
\titleformat{\subsection}{\normalsize\itshape}{\thesubsection.}{0.6em}{}
\titleformat{\subsubsection}{\normalsize\bfseries\itshape}{\thesubsubsection.}{0.6em}{}
\titlespacing*{\section}{0pt}{1.1em}{0.4em}
\titlespacing*{\subsection}{0pt}{0.9em}{0.3em}
\titlespacing*{\subsubsection}{0pt}{0.7em}{0.3em}

\begin{document}

\twocolumn[

\begin{@twocolumnfalse}
\vspace{0.5cm}

{\centering \fontsize{15}{18}\selectfont The Token Efficiency Index: A Peer-Benchmarked Composite Indicator for AI Token Efficiency \par}

\vspace{0.5cm}

{\centering \fontsize{11}{13}\selectfont Caden Wong$^{1,2,*}$, Vikram Das$^{1,*}$, Himanshu Dhami$^{1}$ \par}

\vspace{0.3cm}

{\centering \footnotesize\itshape $^{1}$Yasu, Amsterdam, Netherlands\\ $^{2}$Massachusetts
Institute of Technology, Cambridge, MA, United States\\
\upshape $^{*}$Corresponding authors: cadenw@mit.edu (C. Wong); vikram@yasu.cloud (V. Das) \par}

\vspace{0.7cm} \hrule \vspace{0.5cm}

{\noindent\textbf{Abstract}\par} \vspace{0.3em} {\justifying
As artificial intelligence (AI) adoption accelerates across tech giants, AI-native startups, and non-technical organizations alike, a deceptively simple question remains hard to answer: \textit{is that spending efficient?} AI consumption is priced by tokens, and costs vary by token type (input, output, reasoning) and model type, with usage ranging from a few hundred tokens for simple queries to over a million for multi-step agentic tasks. This variance makes cost comparison, both within and across organizations, difficult without a standardized framework.

We introduce the \textit{Token Efficiency Index} (TEI), a peer-benchmarked \textit{composite indicator} that condenses token spend efficiency into a single 0-100 score. The TEI ingests an organization's AI usage data, independent of the underlying provider, and computes three direction-aware metrics: cache hit rate, cache amortization ratio, and premium model share. These are normalized to a common scale and aggregated via an equal weights composite and a \textit{Benefit-of-the-Doubt} (BoD) \textit{Data Envelopment Analysis} (DEA) model, with a robust order-\textit{m} extension for sparse data. The result is a headline score, a peer percentile, and frontier-gap recommendations with estimated savings. Grounded in established methods from composite indicator and DEA literature, the TEI offers a transparent, interpretable approach to benchmarking AI token efficiency and identifying opportunities to optimize AI spend.\par}
 
\vspace{0.6em} {\small\noindent\textit{Keywords:}\\Token efficiency, Composite indicator, Benefit of the Doubt, Peer benchmarking, Tokenomics, AI cost management, Data Envelopment Analysis, FinOps \par}
 
\vspace{0.5cm} \hrule \vspace{0.7cm}
 
\end{@twocolumnfalse}
]
 
\thispagestyle{firstpage}
 
 
\section{Introduction}
Founded in February 2019 to give cloud practitioners a shared platform for collaboration and best-practice development, the FinOps Foundation has grown from an initial cohort of organizations—including Nationwide, Spotify, Nike, and MIT—into a community representing thousands of companies today \cite{finops2019}. The emerging discipline of \textit{tokenomics} extends that same accountability from cloud infrastructure spend to AI-specific consumption.
 
\subsection{Growth in enterprise AI spend}
The scale of this spending is what makes the question worth asking. Enterprise generative AI spending reached an estimated \$37 billion in 2025, up roughly 3.2x from \$11.5 billion in 2024, and now represents more than 6\% of the entire global software market \cite{menlo2025}. Ramp transaction data shows median business AI spend grew 4x between February 2025 and February 2026 alone, and these companies now devote 15\% of their software budget to AI tools—a share that was negligible only two years earlier \cite{ramp2026}. Unlike traditional, seat-based SaaS costs, AI spend is usage-driven, harder to forecast, and often not owned by a single function \cite{gerber2026}. As a result, this growth has outpaced many organizations' ability to track it.
 
\subsection{Limitations of existing cost-visibility tools}

Cost visibility alone does not resolve this: knowing how much an organization spends on tokens says little about whether that spend reflects efficient usage or simply reflects scale. Distinguishing the two requires efficiency measures that are independent of size and evaluated against what comparable peers achieve.

Yet existing cost-visibility tools are not built for this. They are attribution tools, not efficiency tools: they trace token consumption to a request, team, feature, or customer, showing \textit{what} was spent and often \textit{where}, but not \textit{how efficiently}. What is missing is not more data, but a method for condensing existing usage data into an interpretable score.

\subsection{Towards an interpretable token-efficiency score}

This paper introduces the Token Efficiency Index (TEI), a peer-benchmarked composite indicator that condenses multiple dimensions of token spend efficiency into a single 0-100 score. Rather than reporting caching behavior and model selection as separate figures, the TEI combines them into one score, in the same spirit as composite indicators such as ENERGY STAR have done for physical energy efficiency.

This paper follows the six-stage COOPER framework that Mergoni, Emrouznejad, and De Witte propose for organizing non-parametric frontier studies in their review ``Fifty Years of Data Envelopment Analysis'' \cite{mergoni2025fifty}: concepts and objectives, organizing data, operational models, performance comparison, evaluation, and results and discussion. Section~2 places the TEI within the composite indicator and efficiency scoring literature and clarifies its novelty relative to existing approaches. Section~3 describes the data behind the current implementation. Section~4 details the scoring methodology, including normalization, weighting, and the Benefit of the Doubt approach. Section~5 covers the implementation and evaluates it. Section~6 examines the explainability and sensitivity of the resulting scores, and Section~7 concludes with limitations, future work, and a call to extend this into a broader organizational benchmark.
 
\section{Background and Novelty}
We take the ENERGY STAR score as our primary reference for a robust, peer-benchmarked 0-100 efficiency score.
 
\subsection{Composite indicators and the Energy-Star analogy}
The two scores share several defining features. Both distill a complex, multi-factor efficiency question into one number a non-technical reader can act on. Both are peer benchmarked rather than absolute, comparing each unit explicitly against comparable organizations to give a meaningful measure of relative performance. And both use the same 0-100 scale.
 
The two differ in method, however. ENERGY STAR predicts a building's expected energy use by regressing a measurable output (energy consumed) against building attributes, then scores the ratio of actual to predicted use \cite{arjunan2020energystar}. Token efficiency has no analogous output variable: its metrics are efficiency ratios, not quantities that could be regressed against a predicted output. This rules out a regression-based approach and motivates the alternative scoring method described in Section~2.3.
 
\subsection{Contemporary approaches to efficiency scoring}
A composite indicator measures a concept that no single indicator can capture on its own, combining several sub-indicators into one interpretable measure. Because efficiency depends on size-independent ratios rather than absolute quantities, a composite indicator built from efficiency ratios is a natural fit for the token efficiency problem.
 
Constructing a composite indicator involves a well-established sequence of choices: defining the concept, selecting indicators, normalizing them, and aggregating them. Each step admits several options, and the right choice depends on the data and the problem at hand. The most comprehensive reference is the OECD's \textit{Handbook on Constructing Composite Indicators} \cite{oecd2008}, which catalogs these choices and the trade-offs among them. We also draw on Mazziotta and Pareto \cite{mazziottapareto2013}, which offers a decision roadmap through the same choices, framed around whether a single method can serve all cases or must be tailored to each. Both works informed the design of the TEI; the specific choices we make for normalization, aggregation, and scoring are detailed in Section~4.
 
\subsection{Novelty of the proposed approach}
The TEI's contribution is not a new statistical technique, but the application of an established one to a domain that has no standardized efficiency score yet. First, in the absence of an output variable available, the TEI scores each organization using Benefit of the Doubt (BoD) Data Envelopment Analysis (DEA). BoD derives each organization's weights endogenously and benchmarks it against best observed practice, rather than against an externally imposed standard. Second, because the current peer set is thin and outlier-prone—39 organizations with real usage patterns that vary widely in scale and workload—the TEI adds a robust order-\textit{m} extension of BoD. This benchmarks each organization against random subsets of its peers rather than the single best performer, which reduces the sensitivity to outliers that standard frontier methods tend to show.

Neither choice is unusual. The most recent comprehensive review of the DEA literature identifies BoD-based composite indicators and their robust order-\textit{m} extensions as an active, well-established line of methodological development, not a niche technique borrowed for convenience \cite{mergoni2025fifty}.
 
Together, these make the TEI, to our knowledge, the first peer-benchmarked composite indicator for AI token efficiency.

\section{Data}

Token consumption occurs in two settings. \textit{Internally,} organizations use tokens through AI coding assistants, chatbots, and other developer tools. \textit{Externally,} organizations may deploy AI-powered products built on hyperscaler platforms, where token usage is generated by end users. Extending the TEI to external usage would require production API billing data, which was not available for this study (Section~7.1).

\subsection{Seed data and synthetic population}
Initially, only two organizations' usage logs were available: Yasu's internal data and that of one customer. We used this data to characterize token usage patterns and build a synthetic population for developing and testing the TEI scoring pipeline. We generated the population through Monte Carlo simulation with plausible variation in organizational behavior, and used it solely for methodology development rather than benchmarking.

\subsection{The Tokscale dataset}
During development, we discovered Tokscale \cite{yeo2025tokscale}, an open-source platform that monitors AI coding-assistant usage across more than 40 developer tools and maintains a public, opt-in leaderboard of participating accounts.

\begin{figure}[t]
  \centering
  \includegraphics[width=\columnwidth]{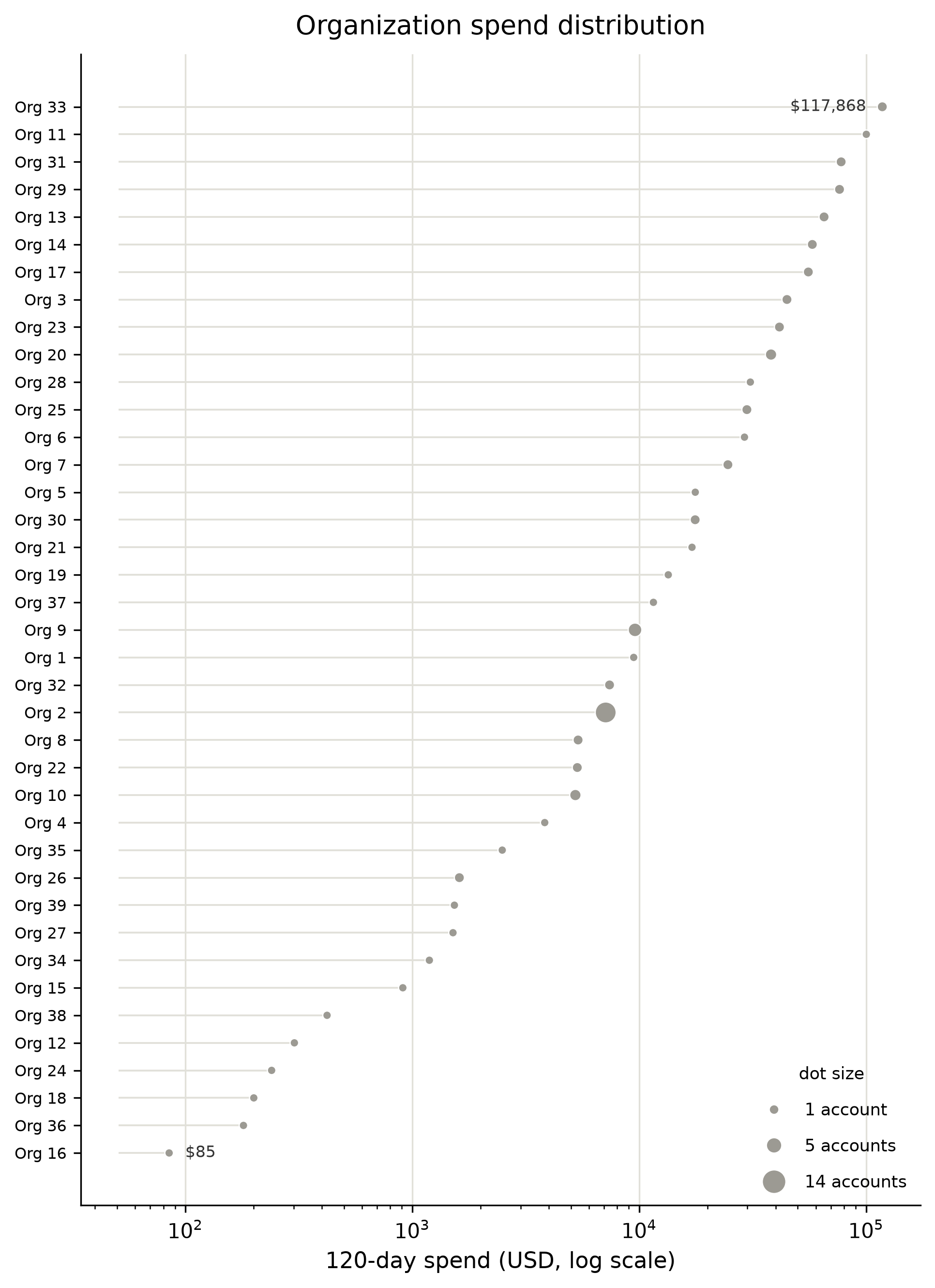}
  \caption{Distribution of 120-day spend across the sample (log scale), sorted from lowest to highest; dot size indicates the number of accounts per organization. Organization identities are anonymized (Section~3.2).}
  \label{fig:org_scale}
\end{figure}

We obtained a snapshot of the leaderboard ending July 16, 2026, containing 78 developer accounts. After excluding a small number of accounts (Section~3.3), the sample comprises 75 accounts grouped into 39 organizations, including individual developers, small teams, technology companies, and research institutions. Organization size, in both number of accounts and 120-day token spend, varies substantially: size ranges from single-account entities (20 of 39 organizations) to a maximum of 14 accounts, and spend varies by roughly three orders of magnitude, from \$85 to \$118{,}000 (Figure~\ref{fig:org_scale}).

Because the dataset captures naturally occurring variation across organizations, it replaced the synthetic population as the empirical basis for the TEI. It also motivated the revised metric set described in Section~3.4, which replaces the earlier prototype metrics. Organization identities are anonymized throughout this paper; we keep each organization's exact usage figures while replacing its name with a fixed ``Org $N$'' label. All results in this paper use this dataset. 

\begin{table}[t]
  \centering
  \small
  \renewcommand{\arraystretch}{1.15}
  \setlength{\tabcolsep}{4pt}
  \begin{tabular}{@{}>{\raggedright\arraybackslash}p{2.1cm} l >{\raggedright\arraybackslash\hyphenpenalty=10000\exhyphenpenalty=10000}p{3.3cm}@{}}
    \toprule
    Metric & Direction & Definition \\
    \midrule
    Cache hit rate & Higher & Cache reads $\div$ (input tokens $+$ cache reads) \\
    Cache amortization ratio & Higher & $\log(\text{cache reads} \div \text{cache writes})$ \\
    Premium model share & Lower & Premium-tier tokens $\div$ total tokens \\
    \bottomrule
  \end{tabular}
  \caption{Token-efficiency metrics used in the TEI. Cache hit rate and premium model share are token-count ratios in $[0,1]$; cache amortization ratio is a log-ratio normalized onto the same scale (Section~4.1). Direction indicates whether higher or lower raw values are more efficient; premium tiers follow the relative pricing method in Section~\ref{sec:premium-tier}.}
  \label{tab:metrics}
\end{table}

\begin{figure*} [t]
    \centering
    \includegraphics[width=0.95\textwidth]{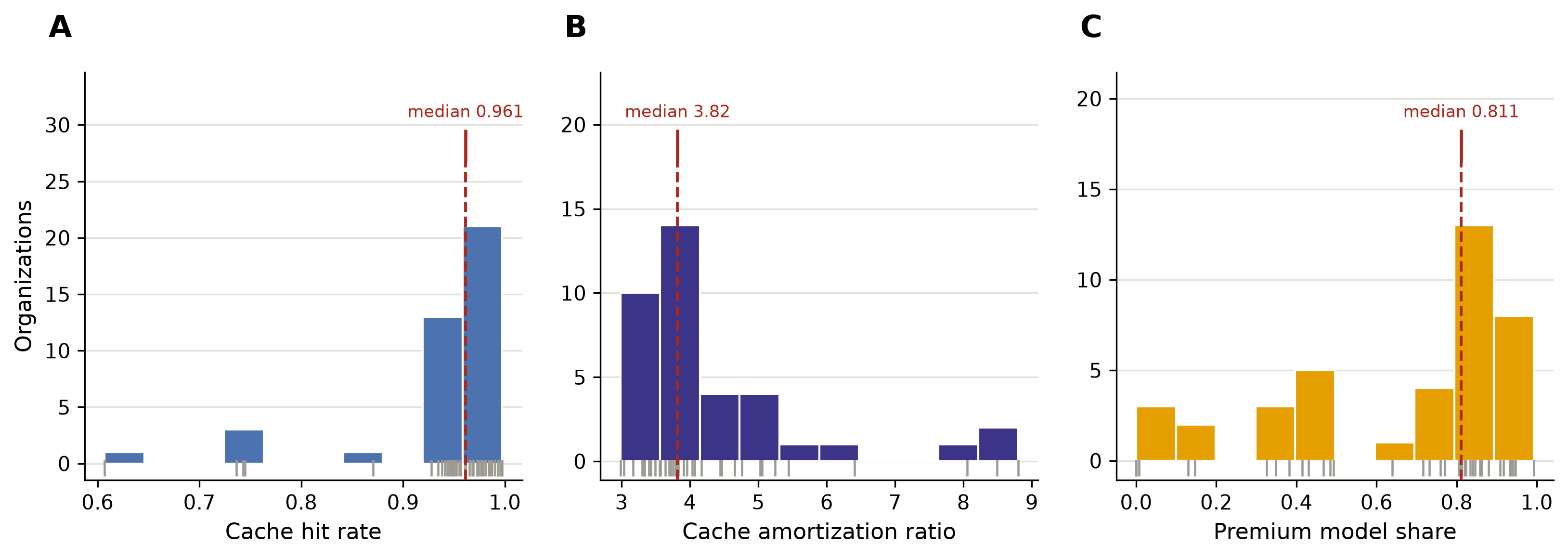}
  \caption{Distribution of TEI input metrics across organizations (A: cache hit rate; B: cache amortization ratio; C: premium model share), with each organization's value marked below the axis and the median labeled. Cache amortization ratio is undefined for two organizations with no cache-write activity ($n=37$; Section~4.2); the other two metrics are defined for all 39.}
  \label{fig:metric_distributions}
\end{figure*}

\subsection{Data cleaning and preprocessing}
Before computing the TEI metrics, we transform the raw usage records into an organization-level dataset. We excluded three accounts without resolvable organizational affiliations, leaving 75 accounts across 39 organizations. Because the source data reports lifetime cumulative totals and account ages vary substantially, we reconstruct metrics from daily usage records over a fixed 120-day observation window ending on the snapshot date, so that differences in account age are not mistaken for differences in efficiency.

We resolve organization affiliations from declared employer information or, when that is unavailable, from other affiliation signals; accounts with no identifiable affiliation are kept as individual organizations. Company and model identifiers are canonicalized through explicit lookup mappings, so that formatting variations and model aliases are aggregated consistently.

All TEI metrics are computed using an aggregate-first approach: we sum token counts and costs across all accounts within an organization before calculating ratios. This avoids the distortions that can arise from averaging account-level ratios for organizations with different numbers of users.

\subsection{Metric selection and validation}

The TEI is built from three metrics representing controllable efficiency levers, summarized in Table~\ref{tab:metrics}: how well an organization reuses context, how efficiently a cache write is recouped through subsequent reads, and how much of its token usage goes toward premium-tier models. Blended cost per million tokens was considered but excluded from the composite score, since it is mainly an outcome of these operational choices rather than an independent efficiency lever. It is instead reported separately alongside the TEI score in Yasu's Cost of AI dashboard.

Given the limited number of organizations, we prioritize a small set of interpretable metrics over a larger set of potentially noisy indicators, consistent with guidance on variable selection in composite indicators \cite{mergoni2025fifty}. Figure~\ref{fig:metric_distributions} shows the distributions of these selected metrics across organizations.

\subsubsection{Defining premium tiers}
\label{sec:premium-tier}

We determine premium tiers from the relative price distribution of observed models rather than a fixed vendor-defined list. For each model with nonzero usage during the 120-day window, we calculate effective price per million tokens, rank models by price, and split them into equal thirds: Premium, Standard, and Budget. In the current snapshot, this produces 68 Premium, 69 Standard, and 69 Budget models across 206 models with observed usage.

This relative approach lets the tier definition adapt as pricing and model availability change. However, Premium represents relative cost within the observed sample rather than an absolute measure of model capability.

Accordingly, this metric treats all premium-tier usage as equally undesirable; it does not distinguish requests that genuinely need a stronger model from those that simply default to one. As a result, it cannot currently account for task difficulty, and may overstate inefficiency when premium models are appropriately chosen for more complex requests. We return to this limitation, and a possible fix, in Section~7.2.

\subsubsection{Assessing metric independence}

Before aggregation, we check whether the selected metrics contain redundant information, using two standard diagnostics: pairwise correlations among the normalized metrics (oriented so that higher values indicate better efficiency), and Cronbach's alpha across the three metrics.

\begin{figure*}[t]
  \centering
  \includegraphics[width=0.8\textwidth]{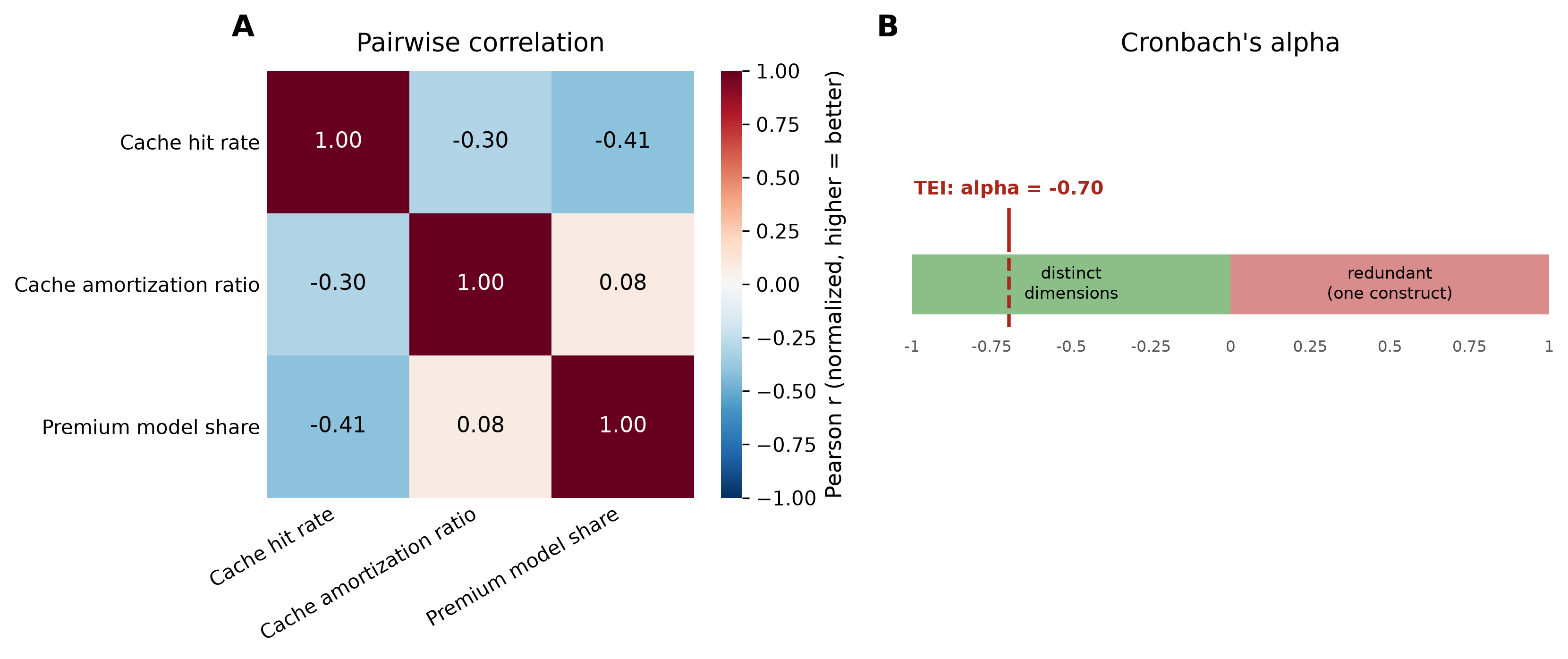}
  \caption{Metric redundancy diagnostics. (A) Pairwise correlations among the normalized metrics (oriented so higher is better); no pair exceeds the conventional $|r|\geq0.7$ threshold. (B) Cronbach's alpha ($\alpha=-0.70$), interpreted in Section~6.1.}
  \label{fig:metric_redundancy}
\end{figure*}

The correlation matrix is shown in Figure~\ref{fig:metric_redundancy}A: cache hit rate and premium model share correlate at $r=-0.41$, cache hit rate and cache amortization ratio at $r=-0.30$, and cache amortization ratio and premium model share at $r=0.08$. Cronbach's alpha for the three normalized metrics is $-0.70$, computed on the 37 organizations with complete coverage of all three metrics (Section~4.2), and is shown alongside the conventional reliability scale in Figure~\ref{fig:metric_redundancy}B. We interpret both diagnostics in Section~6.1.

\section{Scoring Methodology}
This section covers how the TEI turns a matrix of raw metric values into scores: normalization onto a common, direction-aware scale; an equal-weights baseline; the Benefit of the Doubt and robust order-\textit{m} methods at the core of the index; the headline 0-100 mapping and peer percentile; and the frontier-gap recommendations.

\subsection{Normalization}
Because the individual metrics point in different directions (some where higher values are better, others where lower values are better), we first normalize them onto a common, direction-aware scale before aggregating them. So that the final score can be expressed on a 0-100 scale where higher denotes greater efficiency, we orient every metric during normalization so that higher normalized values indicate better performance.

We use min--max normalization, which linearly rescales each metric to the unit
interval $[0, 1]$. For a metric in which higher raw values are better, the
normalized value of observation $i$ is

\begin{equation}
    x_i^{\text{norm}} = \frac{x_i - \min(x)}{\max(x) - \min(x)},
\end{equation}

and for a metric in which lower raw values are better, the direction is
inverted:

\begin{equation}
    x_i^{\text{norm}} = \frac{\max(x) - x_i}{\max(x) - \min(x)},
\end{equation}

where $\min(x)$ and $\max(x)$ are the minimum and maximum observed values of
that metric across all organizations. In both cases the best-performing
organization on a given metric maps to $1$ and the worst to $0$, so a higher
normalized value always denotes better performance.

When a metric has no variation across organizations ($\max(x) = \min(x)$), the
denominator vanishes and the metric cannot tell organizations apart. In
this case, we assign a neutral value of $0.5$ to every observation rather than
divide by zero, so that a non-discriminating metric neither rewards nor
penalizes any organization.

One metric gets an additional transformation before this normalization step. Cache amortization ratio is defined as $\log(\text{cache reads} \div \text{cache writes})$ (Table~\ref{tab:metrics}) rather than the raw read:write ratio. Across the sample, that raw ratio spans roughly three orders of magnitude and compresses sharply as organizations amortize their cache writes over more reads, leaving little room to distinguish organizations at the high end of the distribution. Taking the natural log spreads out this compressed range, recovering discriminating power among the organizations that reuse cache most heavily, before we min--max normalize the resulting value like the other two metrics.

In the current sample, two organizations lack cache amortization ratio because they have no cache-write activity; every other organization has complete metric coverage. We handle missing metrics by scoring each organization using only its available metrics, while still letting it serve as a peer in other organizations' reference sets. We apply this rule consistently across all three scoring methods in this section, including the Benefit of the Doubt and robust order-$m$ approaches described below.

\begin{figure*}[t]
  \centering
  \includegraphics[width=0.9\textwidth]{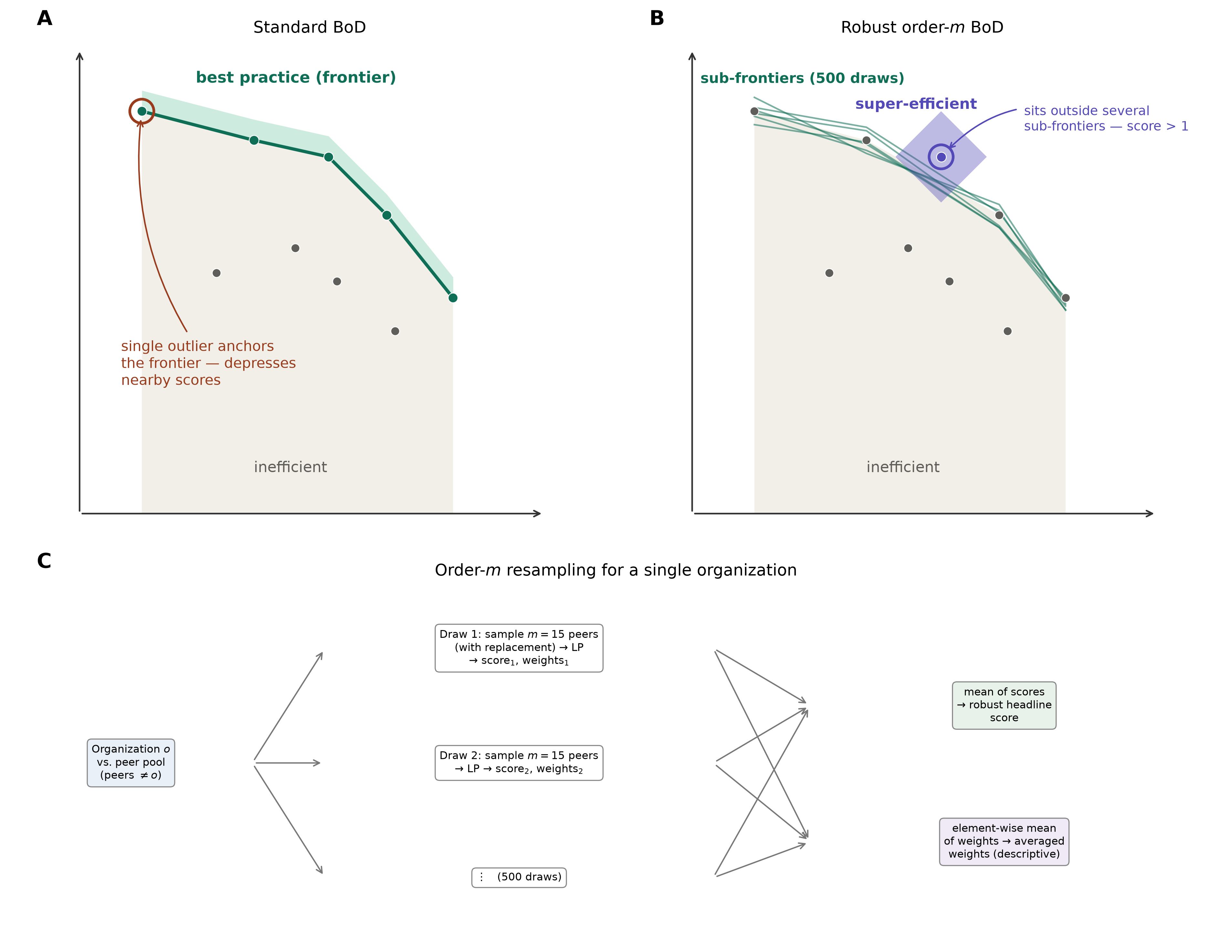}
  \caption{The Benefit of the Doubt frontier and how robust order-$m$ resamples it (Section~4.3--4.4). (A) Standard BoD: a single best-practice frontier (solid line) envelopes the peer cloud, anchored by one outlier (circled). (B) Robust order-$m$ BoD: many sub-frontiers (faint lines), each from a random $m$-peer subsample; an organization lying outside several at once (circled) is super-efficient. Panels A--B use two illustrative dimensions for visualization; the actual TEI frontier spans all three metrics. (C) The order-$m$ resampling procedure for a single organization.}
  \label{fig:bod_method}
\end{figure*}
\subsection{Equal-weights composite (baseline)}
Following standard composite indicator practice \cite{oecd2008}, we first build an equal-weights composite to serve as a baseline. Each normalized metric gets equal weight, and we combine the three metrics into a single score on $[0,1]$ using an arithmetic mean. When a metric has no variation across organizations, it contributes the neutral value of 0.5 set during normalization.

The equal-weights composite is intentionally simple, and its role is diagnostic. If our more sophisticated scoring methods rank organizations almost identically to equal weighting, the added complexity is hard to justify. If the methods rank organizations differently, that divergence shows exactly what the more sophisticated methods contribute. Equal weighting thus gives us a transparent reference point for judging the Benefit of the Doubt methods introduced below.

\subsection{Benefit of the Doubt scoring}
Our primary scoring method is the Benefit of the Doubt (BoD) approach, a DEA-based weighting scheme first introduced by Melyn and Moesen \cite{melynmoesen1991} for benchmarking macroeconomic performance and later formalized for composite-indicator construction by Cherchye, Moesen, Rogge, and Van Puyenbroeck \cite{cherchye_bod}. The method has since been applied in practice: Moesen and Cherchye \cite{moesen1998limep} used the same formulation to build the Limited Information Macroeconomic Performance (LIMEP) index for OECD countries, for the same reason that motivates us here—no principled basis exists for setting fixed weights across competing objectives in advance.

BoD sidesteps this by not imposing weights externally. Instead, for each organization it solves a separate linear program that picks the weights
most favorable to that organization, subject to the requirement that no
organization scores above one under those same weights:
\begin{equation}
    \mathrm{CI}_o \;=\; \max_{w \geq 0}\; \sum_{i} w_i\, x_{io}
    \quad\text{s.t.}\quad
    \sum_{i} w_i\, x_{ij} \leq 1 \;\;\; \forall\, j,
\end{equation}

where $x_{io}$ is the normalized value of metric $i$ for the scored organization
$o$, and the constraints range over every organization $j$ in the reference set (we handle missing metrics per Section~4.2 for both the scored organization and its peers).
Because each organization appears in its own reference set, its score is bounded
above by one; BoD scores therefore lie in $(0, 1]$, with a score of one denoting
an organization on the best-practice frontier.

\subsubsection{Interpreting Benefit of the Doubt weights}
The appeal of this construction is that every organization is evaluated under the weighting most favorable to it. An organization is not penalized for performing relatively poorly on a dimension it does not emphasize, as long as it performs well on the dimensions it does; a low score arises only when it is outperformed under every admissible weighting. The resulting endogenous weights are themselves informative: they reveal the metrics on which each organization has a comparative advantage.

This flexibility comes with two well-known drawbacks. When weights only need to be non-negative, many organizations reach the maximum score of one simply by putting all their weight on the one metric where they excel, leaving the other metrics almost no weight \cite{cherchye_bod} (in our sample, 7 of 39 organizations achieve a perfect score, several by placing nearly all weight on a single metric). In addition, because standard BoD benchmarks every organization against the full-sample frontier, it can be highly sensitive to outliers. These limitations motivate the robust, order-$m$ formulation described next.

\subsection{Robust order-\textit{m} Benefit of the Doubt}
To reduce this outlier sensitivity, we use a robust order-$m$ approach, following Cazals, Florens, and Simar's non-parametric order-$m$ frontier estimator \cite{cazals2002}. Rather than benchmarking against the full population frontier, we compare each organization against many random subsamples of $m$ peers, drawn independently with replacement from organizations that share its available metrics (Section~4.2), and average the resulting scores. Because no single outlier appears in every subsample, its influence on any one score is diluted, leaving the results far less sensitive to extreme observations than standard BoD (Figure~\ref{fig:bod_method}A,B).

We draw $m = 15$ peers per subsample and average over 500 draws. These values balance two competing pressures: a smaller $m$ gives greater robustness but coarser discrimination, while more draws reduce Monte Carlo noise at the cost of computation.

Each of the 500 draws solves its own linear program and produces a corresponding per-metric weight vector. Alongside the headline score, the pipeline reports the element-wise average of these 500 vectors for each organization, computed over draws whose optimization problem solved successfully. This average is not an optimal weight vector, since each draw's weights are optimal only for its own sampled reference set. Instead, it serves as a descriptive summary of which metrics an organization tends to emphasize across the resampling procedure, providing a useful complement to the headline score. Figure~\ref{fig:bod_method}C illustrates this procedure end to end for a given organization.

\subsubsection{Super-efficiency and parameter selection}
A distinctive consequence of the order-\textit{m} approach is \emph{super-efficiency}. Because each organization is compared only against a subsample of its peers—typically excluding the single strongest performer—a sufficiently strong organization can outperform its entire reference set and receive an average score above one. Unlike standard BoD, which caps all scores at one, robust BoD therefore preserves discrimination among top performers. In rare cases, however, a subsample may leave a metric effectively uncontested, producing an unbounded score; we cap the contribution of each draw at two to prevent this.

\subsection{Headline score and peer percentile}
For interpretability, we map the robust BoD score to a $0$--$100$ headline score, with the best-practice frontier (a robust BoD score of one) mapped to $100$. Super-efficient organizations also receive a headline score of $100$; to preserve distinctions among them, we additionally report the raw robust BoD score alongside the headline value. We choose this absolute, frontier-anchored mapping over a population min--max rescaling so that the score keeps a fixed interpretation regardless of which peers happen to be included in the comparison.

Each organization's peer percentile is the proportion of peers scoring at or below its headline score, expressed as a percentage. Normalization does not eliminate structural differences between organizations, however: smaller or newer organizations may simply not yet match the cache-hit rate or model-tier mix of larger, established peers. The pipeline addresses this by supporting percentiles computed within peer cohorts (e.g., by organization size or industry), not just against the full population. The current release still treats the sample as a single cohort, because we did not collect size and industry attributes. Cohort segmentation is already implemented and will switch on once these attributes are gathered.

\subsection{Frontier-gap recommendations}
Because BoD scores each organization relative to a frontier of its peers, the same framework can also flag improvement opportunities. For each metric, we define a frontier target using a high-performing percentile of organizations rather than the single best performer, which reduces sensitivity to outliers. We then evaluate an organization's gap to this target, and metrics that fall beyond a set threshold generate recommendations.

Each recommendation is linked to an estimated cost saving, based on the relevant pricing lever—for example, the cost difference between premium and economy model tiers, or discounted prompt-cache reads. Cache amortization ratio has no such lever: it is a log-ratio, not a bounded share of spend, so converting its gap into a dollar figure would require assumptions that break down given how widely read/write volumes vary across organizations (by two to three orders of magnitude). For this metric, recommendations instead report a qualitative LOW/MEDIUM/HIGH severity tier, based on the organization's percentile rank among peers with a gap on this metric. We rank dollar-estimated recommendations by estimated financial impact; severity-tiered recommendations are reported alongside them, not folded into that ranking. Because these estimates come from aggregate usage statistics rather than per-call API data, they should be read as directional indicators of potential savings, not precise forecasts.

\subsection{Aggregation form: additive versus geometric}

Section 4.2 introduced the equal-weights composite as a transparent baseline, combining the three normalized metrics using an unweighted arithmetic mean. The pipeline also supports a second aggregation rule for this baseline: replacing the arithmetic mean with the geometric mean while using the same normalized inputs. Given the normalized metric values $x_{io}$ for organization $o$, the two forms are

\begin{equation}
    \mathrm{CI}_o^{\mathrm{add}} \;=\; \frac{1}{n}\sum_i x_{io},
    \qquad
    \mathrm{CI}_o^{\mathrm{geo}} \;=\; \left(\prod_i x_{io}\right)^{1/n},
\end{equation}

where $n$ is the number of metrics.

The two forms differ in how much one metric can compensate for another. The arithmetic mean lets a high score on one metric fully offset a low score on another, while the geometric mean does not: a metric value close to zero pulls the entire product toward zero regardless of performance elsewhere. Neither assumption is inherently better. Rather than picking one, we report both as a second axis of sensitivity analysis alongside the equal-weights-versus-BoD comparison introduced earlier. Where the two forms produce similar rankings, the choice of aggregation rule has little practical consequence. Where they diverge substantially, the disagreement itself is informative: it shows that an organization's apparent efficiency under the arithmetic mean depends on compensating for a genuinely weak metric, rather than performing consistently well across all metrics.

We currently implement this comparison only for the equal-weights baseline, not for BoD or robust order-$m$, whose weights are optimization outputs rather than fixed inputs to aggregate (Section~7.1); we discuss extending it to these adaptive schemes in Section~7.2.

The geometric mean also requires strictly positive metric values, since a single zero collapses the entire product. Non-discriminating metrics never produce a zero (they receive the neutral value $0.5$), but discriminating metrics assign the worst-performing organization exactly zero under min--max scaling. The pipeline floors all normalized values at a small constant ($10^{-9}$) before taking logarithms, which keeps the geometric mean well defined with negligible effect on other scores.

\begin{figure}[t]
  \centering
  \includegraphics[width=\columnwidth]{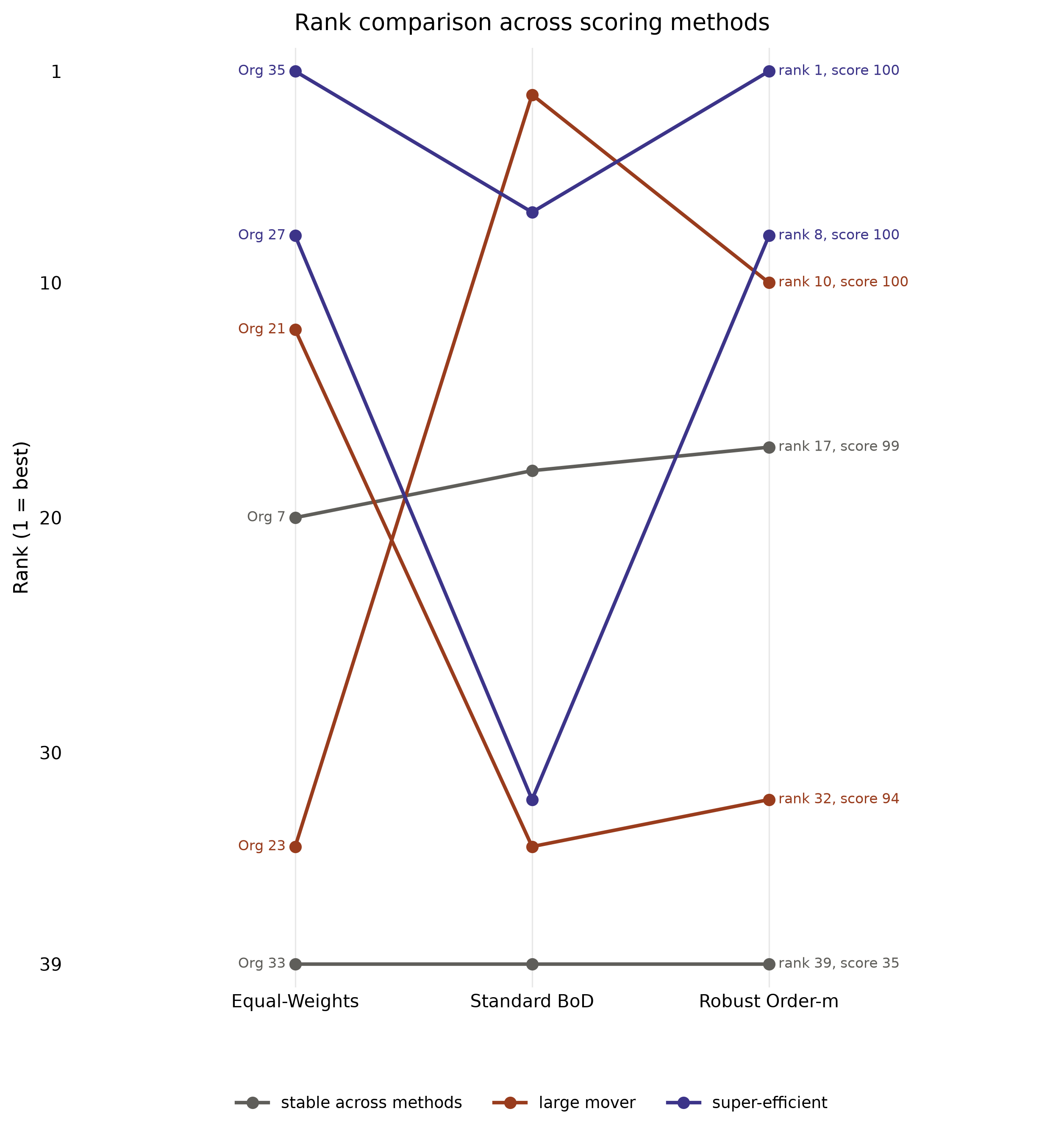}
  \caption{Rank comparison across equal-weights, standard BoD, and robust
  order-$m$ scoring for a representative subset of six organizations (full
  population $n=39$): two rank-stable, two large movers, and two
  super-efficient under robust order-$m$. Organization identities are
  anonymized (Section~3.2); the rightmost column reports each
  organization's robust order-$m$ rank and 0--100 headline score.}
  \label{fig:rank_comparison}
\end{figure}
\section{Implementation}

We implement the TEI as a configuration-driven scoring engine in Python. A
single configuration file specifies every run-time choice---including the
metric set and optimization direction for each metric, the normalization
method, the aggregation rule, the order-$m$ parameters, and the recommendation
levers---so that changing the methodology requires only a configuration edit
rather than a code change. The engine builds on standard scientific Python
components, including schema-validated data ingestion that rejects malformed
usage exports before they reach the scorer, vectorized normalization, and
Benefit of the Doubt linear programs solved using the HiGHS solver. Every
reported quantity is fully reproducible from fixed random seeds rather than
persisted intermediate results: both the order-$m$ resampling procedure and
the sensitivity analyses draw from seeded random number generators, so
re-running the pipeline reproduces the reported results exactly.

\subsection{Evaluation approach}
This release evaluates the engine on the real usage dataset described in Section~3, with
the three efficiency metrics computed directly from raw usage records and
passed unchanged to the scoring pipeline. Evaluation proceeds along two
tracks: population-level behavior on the full dataset of 39 organizations
(Section 5.2), and a targeted perturbation analysis that checks whether the engine
responds to controlled input changes in the expected direction and magnitude
(Section 5.3).

We judge correctness against the mechanical properties the engine is
designed to satisfy: scores fall within expected ranges, discriminating and
non-discriminating metrics are handled appropriately, and the ranking
guarantees of Section 4 hold---for instance, that an organization is never
penalized for a metric it does not emphasize. Because two organizations are
missing one metric (cache amortization ratio; Section~4.2) but all three
metrics otherwise vary across the sample, this evaluation exercises the
missing-metric skip described in Section~4.2, in addition to the guard
conditions we validate separately through targeted tests of the
metric-construction stage (the zero-variance neutral value and the
discriminating-column filter).

\subsection{Ranking behavior across scoring methods}
The three scoring methods agree in broad strokes while
differing in instructive ways. Rank correlations are all
positive: Spearman $\rho=0.14$
between equal weights and standard BoD, 0.77 between standard and robust
BoD, and 0.54 between equal weights and robust BoD. Robust BoD tracks
equal weights far more closely than standard BoD does (0.54 versus 0.14). This is
reassuring rather than surprising: resampling the frontier tempers the
aggressive single-metric reweighting that lets standard BoD reward an
organization for excelling on just one dimension, pulling rankings back
toward a more balanced reading while still crediting genuine strength.

Figure~\ref{fig:rank_comparison} makes this concrete for a representative
subset. Most organizations move little across the three methods; the
informative cases are the large movers. Org 27 is the clearest: equal weights
already ranks it eighth, but standard BoD drops it to rank 32, before
robust order-$m$ recovers it to eighth again (raw score 1.03, just past the
super-efficiency threshold). Org 23 shows the complementary case: equal weights
ranks it a poor 34th, unable to reward its near-frontier cache hit rate once it
is averaged against a weak premium model share, whereas both BoD variants
recognize that strength by concentrating weight on it, placing it second
under standard BoD and tenth under robust order-$m$.

Beyond ranking, the same frontier structure also drives the recommendation
engine (Section~4.6): every one of the 39 organizations receives at least one
recommendation, reflecting how tightly the best-practice frontier is defined
by the sample's strongest performers. The most common opportunities involve
shifting suitable workloads to lower-cost model tiers (34 organizations),
improving cache write reuse (29 organizations), and improving prompt caching
itself (7 organizations).

\subsection{Perturbation validation}
To test whether the scorer responds correctly to controlled changes, we generate synthetic ``twins'' for each of the 39 real organizations and each of the 3 scored metrics. Each twin uses one of three perturbation magnitudes (small, medium, large) in one of two directions (improve, degrade), holding the organization's other metrics fixed. A twin's peers are always the other 38 real organizations, never other twins, so that a twin's baseline and its perturbed score are compared against an identical peer pool.

Of the $39\times3\times3\times2=702$ possible perturbations, 12 are excluded because the underlying organization is missing that metric entirely (the two organizations missing cache amortization ratio; Section~4.2), and 6 more are boundary-blocked (an organization already at the boundary of a bounded metric cannot be perturbed further in that direction), leaving 684 testable cases. Of these, 665 (97.2\%) produce a measurable score change in the expected direction, and 19 (2.8\%) show no detectable change rather than an incorrect one; the engine produces zero wrong-direction movements across the full test. This pattern holds consistently across metrics: cache amortization ratio 97.3\% (n=222), cache hit rate 97.4\% (n=234), and premium model share 96.9\% (n=228).

To check that the scorer responds proportionally to perturbation size, we compute the Spearman correlation between perturbation magnitude and the resulting absolute score change, separately for each metric: $\rho=0.62$ for cache hit rate, $\rho=0.68$ for premium model share, and $\rho=0.28$ for cache amortization ratio (all $p<10^{-4}$). As a complementary, informational check within individual organizations, we also compute the Pearson $R^2$ across each organization's three perturbation magnitudes for a given metric and direction; the median across 173 such series is 0.99, showing that score changes closely track perturbation magnitude once an organization's own weights and peer reference set are held fixed.

Together, these results show that the scoring engine responds consistently, proportionally, and in the expected direction to controlled changes in the input metrics. They validate the implementation of the scoring procedure rather than the absolute correctness of any organization's score, since BoD provides no independent ground truth against which such scores can be evaluated (Section~2.1).

\section{Explainability and Sensitivity}

A score is only useful if a reader can trust it and understand what it means. That requires answering two questions beyond how the score is computed: can a
given score be explained, and how sensitive is it to the methodological choices
behind it? This section addresses both. We first revisit the internal consistency diagnostics for the three metrics (Section 6.1), then quantify each metric's contribution to the score and its stability under alternative construction choices (Section 6.2), and finally break down an individual organization's score into an organization-specific explanation (Section 6.3).

\subsection{Internal consistency and redundancy}

Section~3.4.2 reported two diagnostics for checking whether the three metrics capture distinct information rather than repeatedly measuring the same underlying construct: pairwise correlations among the normalized metrics (Figure~\ref{fig:metric_redundancy}A) and Cronbach's alpha (Figure~\ref{fig:metric_redundancy}B). Neither result points to meaningful redundancy: no pairwise correlation exceeds the commonly used $|r|\geq0.7$ threshold, and the strongest association, between cache hit rate and premium model share, falls well short of it.

Cronbach's alpha ($\alpha=-0.70$) needs a different interpretation from its more familiar use in reliability analysis. There, a high alpha is desirable because multiple variables are meant to measure the same latent construct. The TEI's metrics are designed for the opposite purpose: caching behavior, cache-write amortization, and model-tier selection represent distinct efficiency levers rather than interchangeable measurements of a single concept. A high alpha would therefore suggest unwanted redundancy, whereas a low—or, in this case, negative—alpha indicates that the metrics capture complementary aspects of organizational behavior rather than repeatedly measuring the same phenomenon.

Taken together, these diagnostics support keeping all three metrics in the composite rather than reducing dimensionality.
\begin{figure}[t]
  \centering
  \includegraphics[width=\columnwidth]{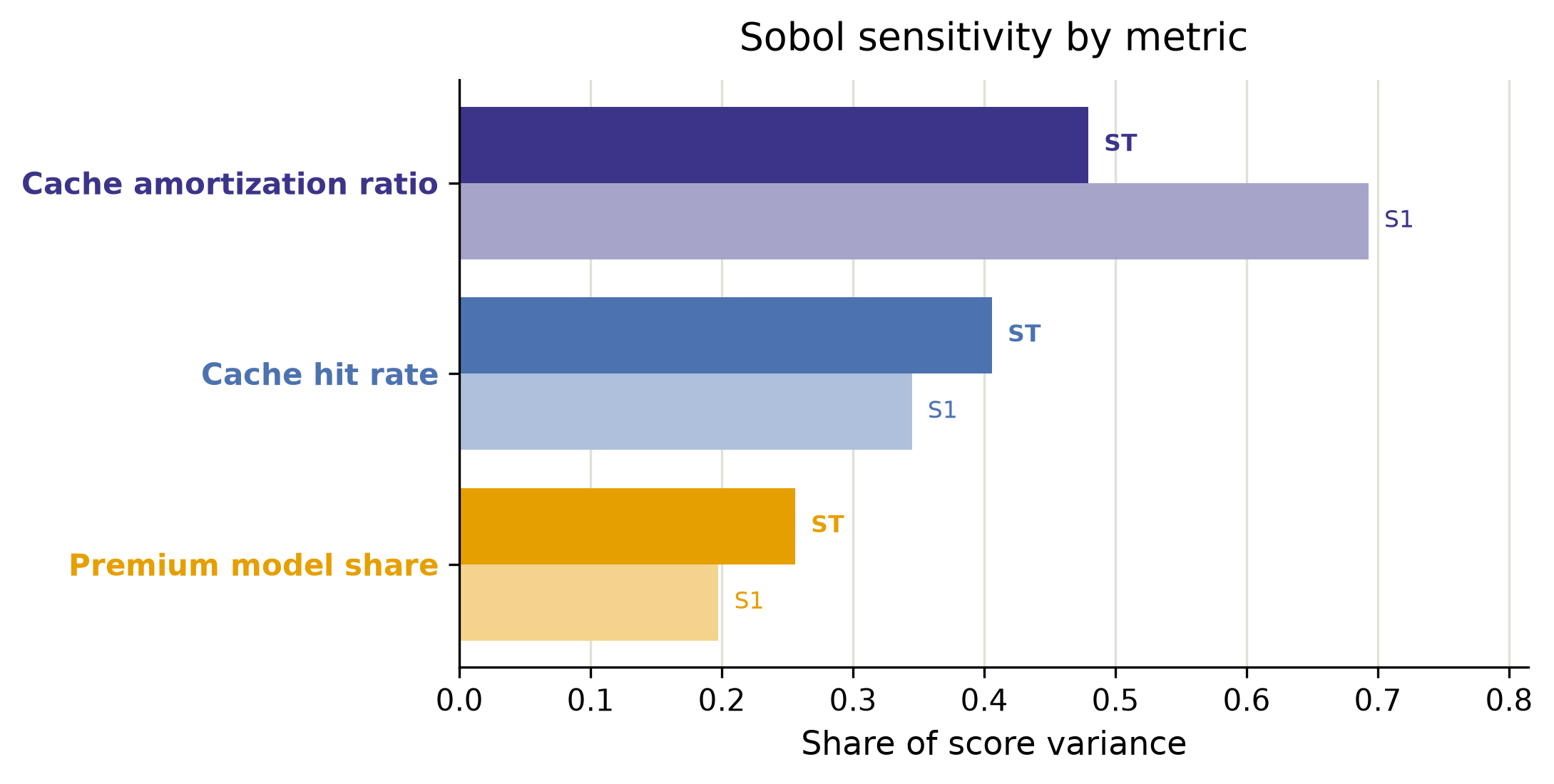}
  \caption{First-order ($S_1$) and total ($S_T$) Sobol sensitivity indices per metric for the robust order-$m$ headline score, discussed in Section~6.2.}
  \label{fig:sobol}
\end{figure}
\subsection{Sensitivity analysis}

We evaluated sensitivity along two complementary axes: which metrics drive the
headline score, and how stable scores remain under alternative
construction choices.

For the first, we compute first-order ($S_1$) and total ($S_T$) Sobol indices
using the robust order-$m$ headline score (Figure~\ref{fig:sobol}). Every metric contributes
meaningfully to score variation: all total indices exceed the common
factor-fixing threshold of 0.05, backing up the redundancy analysis in Section~6.1
that each metric carries distinct information. Under robust order-$m$, total
effects range from 0.26 to 0.48: cache amortization ratio is the largest contributor
(0.48), followed by cache hit rate (0.41), with premium model share the smallest
contributor (0.26). Under the equal-weights composite, contributions are more evenly balanced (0.25--0.46), and premium model share becomes the largest contributor, since equal weighting cannot concentrate influence on whichever metric most strongly tells organizations apart.

One result calls for caution. Under robust order-$m$, cache amortization ratio's
first-order index (0.69) exceeds its own total index (0.48)---an ordering the
Sobol decomposition should not produce under independent inputs. This reflects
its moderate correlation with cache hit rate ($r=-0.30$; Figure~\ref{fig:metric_redundancy}A). Because the Sobol estimator assumes the inputs are independent—it samples each metric on its own, from its own distribution—correlated inputs like these can distort the
standard first-order estimator in exactly this way. We should therefore read the resulting indices
as approximate rather than exact variance
decompositions.

To assess methodological robustness, we also perturb the normalization bounds
and aggregation choices used during construction and measure the resulting
variation in each organization's score. Across these perturbations, score bands
remain narrow, showing that the robust order-$m$ formulation is not only
resistant to outlying peer organizations but also stable under reasonable
variations in construction choices.

\subsection{Explaining an individual score}

Because BoD assigns each organization its own endogenous weights, every score has an organization-specific explanation rather than a single population-wide formula. Consider Org 33, which receives a headline score of 34.7 (raw robust BoD score 0.347) and a peer percentile of 2.6---the lowest values in the sample. The difference between these two statistics is telling: a score in the mid-30s might look moderate on a 0--100 scale, but the peer percentile shows that Org 33 performs below every other organization in the cohort. This happens because the current sample is concentrated near the frontier: 36 of 39 organizations score above 90, and 13 are super-efficient. In such a distribution, the headline score alone compresses distinctions among lower-performing organizations, whereas the percentile preserves their relative position.

Org 33's averaged order-$m$ weights show that its score is driven mainly by cache hit rate (0.98), with smaller contributions from cache amortization ratio (0.23) and premium model share (0.07). Even under this favorable weighting, though, its cache hit rate (73.6\%) remains far below the frontier reference (99.3\%), leaving the organization without a dimension on which it is competitive with the strongest peers.

The recommendation layer translates this gap into potential improvement opportunities. Org 33's largest identified opportunity is premium model share: its premium-tier usage is 90.8\% compared with a frontier reference of roughly 14.3\%, for an estimated directional saving of about \$72{,}150. Improving cache hit rate is a second opportunity, with 73.6\% usage compared with a frontier near 99.3\% and an estimated directional saving of about \$27{,}260. Cache amortization ratio is also flagged as a HIGH-severity gap; because this metric is a log-ratio rather than a bounded share of spend, it is reported as a severity tier rather than converted into a dollar estimate (Section~4.6).

Together, the averaged weights and frontier-gap recommendations show both why Org 33 gets its score and where the largest opportunities for improvement lie.

\section{Conclusion}

We have presented the Token Efficiency Index (TEI), a peer-benchmarked
composite indicator that condenses three direction-aware efficiency metrics into
a single, interpretable $0$--$100$ score. Built for a setting with no natural
output variable and no principled basis for setting fixed metric weights, the
TEI combines the Benefit of the Doubt (BoD) approach with a robust order-$m$
extension that benchmarks organizations against repeated random peer subsets
rather than a single best-performing frontier. The result is a scoring
framework that is adaptive, resistant to outliers, and accompanied by
organization-specific explanations and actionable recommendations.

\subsection{Limitations}

The current release has five main limitations, each discussed in the
relevant methodological section and summarized here for completeness.

First, the averaged order-$m$ weight vectors are descriptive summaries across
the resampling procedure rather than the solution to a single optimization
problem. Second, we currently implement the additive-versus-geometric sensitivity analysis
only for the equal-weights baseline, not for the
endogenous weighting schemes used by BoD and robust order-$m$. Third, peer
percentiles are currently computed against the full sample because
no organization attributes are available for cohort-based benchmarking,
although the pipeline already supports cohort-specific comparisons. Fourth, the
recommendation engine reports directional estimates of potential cost savings
drawn from aggregate usage data, and one metric (cache amortization ratio) is
reported as a qualitative severity tier rather than a dollar estimate at all
(Section~4.6); producing precise savings estimates across all three metrics
would require per-request billing and usage records. Finally, the present evaluation
covers only internally generated token usage from developer tools. Extending
the TEI to externally facing AI applications would require production API
billing data capturing end-user token consumption, which was not available for
this study.

These limitations do not reflect shortcomings of the underlying methodology,
but rather the current availability of data and the scope of the present
implementation.

\subsection{Future work}
 
Several extensions follow naturally from this work. The additive-versus-geometric comparison could be extended to the endogenous weighting schemes of BoD and robust order-$m$, following Verbunt and Rogge~\cite{verbunt2018geometric}.
 
The metrics themselves also offer room for refinement. In particular,
the current premium-model-share metric treats all premium-model usage
as equally undesirable, and does not yet adjust for task difficulty (Section~3.4). In practice, some requests genuinely need more
capable models, whereas others could be served at lower cost without sacrificing
quality. Recent work on LLM routing and model cascades addresses exactly this problem by learning which requests need
stronger models and which can be handled by cheaper alternatives~\cite{ong2024routellm}. A natural extension
would replace premium-model share with a model-misallocation rate:
the share of premium-model usage that could have been served by a less
expensive model while maintaining acceptable quality. Measuring this requires a per-request difficulty or quality signal that current usage exports do not carry.
 
More broadly, the TEI shows that composite-indicator methodology can be adapted to operational AI cost benchmarking. As AI usage continues to expand across organizations, we hope this framework offers a practical and extensible foundation for measuring, comparing, and ultimately improving token efficiency.
 
\subsection{Toward a broader benchmark}
 
The methodology described here is already available today through Yasu's Cost of AI dashboard (Section~3.4). But a peer-benchmarked score is only as good as its peer set, and the current release draws on a thin cohort of 39 organizations. Organization-wide usage, negotiated pricing, and a measure of the value AI systems create would each sharpen future versions of the index—none of which are visible from public data alone.
 
We see this release as a starting point rather than a finished benchmark. Organizations interested in contributing anonymized usage data as described in Section~3.2, or piloting the TEI on their own AI usage, are encouraged to reach out via the author email addresses listed on the first page.

\section*{Author Credit Statement}
Caden Wong: Conceptualization, Methodology, Software, Formal analysis, Writing – Original Draft. Vikram Das: Writing – Review \& Editing. Himanshu Dhami: Data Curation, Writing – Review \& Editing.

\printbibliography

@misc{finops2019,
  author       = {{FinOps Foundation}},
  title        = {About the FinOps Foundation},
  howpublished = {\url{https://www.finops.org/about/}},
  note         = {Accessed 2026},
  year         = {2026}
}

@misc{menlo2025,
  author       = {Tully, Tim and Redfern, Joff and Das, Deedy and Xiao, Derek},
  title        = {2025: The State of Generative AI in the Enterprise},
  howpublished = {\url{https://menlovc.com/perspective/2025-the-state-of-generative-ai-in-the-enterprise/}},
  organization = {Menlo Ventures},
  year         = {2025},
  month        = dec
}

@misc{gerber2026,
  author       = {Gerber, Jeff},
  title        = {What Does Enterprise AI Actually Cost? A Finance Leader's Breakdown},
  howpublished = {\url{https://suplari.com/blog/what-does-enterprise-ai-actually-cost}},
  organization = {Suplari},
  year         = {2026},
  month        = jun
}

@misc{ramp2026,
  author       = {{Ramp}},
  title        = {3 Steps to Manage AI Spend},
  howpublished = {\url{https://ramp.com/3-steps-to-manage-ai-spend}},
  note         = {Accessed 2026}
}

@article{arjunan2020energystar,
  author  = {Arjunan, Pandarasamy and Poolla, Kameshwar and Miller, Clayton},
  title   = {EnergyStar++: Towards more accurate and explanatory building energy benchmarking},
  journal = {Applied Energy},
  volume  = {276},
  pages   = {115413},
  year    = {2020},
  doi     = {10.1016/j.apenergy.2020.115413}
}

@techreport{melynmoesen1991,
  author      = {Melyn, Wim and Moesen, Willem},
  title       = {Towards a Synthetic Indicator of Macroeconomic Performance: Unequal Weighting when Limited Information is Available},
  institution = {Centrum voor Economische Studi{\"e}n (CES), KU Leuven},
  type        = {Public Economics Research Paper},
  number      = {17},
  year        = {1991}
}

@article{mazziottapareto2013,
  author  = {Mazziotta, Matteo and Pareto, Adriano},
  title   = {Methods for Constructing Composite Indices: One for All or All for One?},
  journal = {Rivista Italiana di Economia Demografia e Statistica},
  volume  = {67},
  number  = {2},
  pages   = {67--80},
  year    = {2013}
}

@techreport{oecd2008,
  author      = {{OECD}},
  title       = {Handbook on Constructing Composite Indicators: Methodology and User Guide},
  institution = {OECD Publishing},
  year        = {2008}
}

@article{mergoni2025fifty,
  author  = {Mergoni, Anna and Emrouznejad, Ali and De Witte, Kristof},
  title   = {Fifty years of Data Envelopment Analysis},
  journal = {European Journal of Operational Research},
  volume  = {326},
  number  = {2},
  pages   = {389--412},
  year    = {2025},
  doi     = {10.1016/j.ejor.2024.12.049}
}

@article{cherchye_bod,
  author  = {Cherchye, Laurens and Moesen, Willem and Rogge, Nicky and Van Puyenbroeck, Tom},
  title   = {An Introduction to 'Benefit of the Doubt' Composite Indicators},
  journal = {Social Indicators Research},
  volume  = {82},
  number  = {1},
  pages   = {111--145},
  year    = {2007},
  doi     = {10.1007/s11205-006-9029-7}
}

@techreport{moesen1998limep,
  author      = {Moesen, Willem and Cherchye, Laurens},
  title       = {The Macroeconomic Performance of Nations: Measurement and Perception},
  institution = {KU Leuven},
  type        = {Discussion Paper Series},
  number      = {DPS 98.22},
  year        = {1998}
}

@article{cazals2002,
  author  = {Cazals, Catherine and Florens, Jean-Pierre and Simar, L{\'e}opold},
  title   = {Nonparametric frontier estimation: a robust approach},
  journal = {Journal of Econometrics},
  volume  = {106},
  number  = {1},
  pages   = {1--25},
  year    = {2002},
  doi     = {10.1016/S0304-4076(01)00080-X}
}

@article{verbunt2018geometric,
  author  = {Verbunt, Paulien and Rogge, Nicky},
  title   = {Geometric composite indicators with compromise Benefit-of-the-Doubt weights},
  journal = {European Journal of Operational Research},
  volume  = {264},
  number  = {1},
  pages   = {388--401},
  year    = {2018},
  doi     = {10.1016/j.ejor.2017.06.061}
}

@article{ong2024routellm,
  author  = {Ong, Isaac and Almahairi, Amjad and Wu, Vincent and Chiang, Wei-Lin and Wu, Tianhao and Gonzalez, Joseph E. and Kadous, M. Waleed and Stoica, Ion},
  title   = {RouteLLM: Learning to Route LLMs with Preference Data},
  journal = {arXiv preprint arXiv:2406.18665},
  year    = {2024}
}

@misc{yeo2025tokscale,
  author       = {Yeo, Junho},
  title        = {Tokscale: A {CLI} Tool and Dashboard for Tracking Token Usage and Costs Across {AI} Coding Agents},
  year         = {2025},
  howpublished = {\url{https://github.com/junhoyeo/tokscale}},
  note         = {GitHub repository; MIT license; accessed 2026-07-16}
}

\end{document}